\documentstyle[12pt,epsfig]{article}
\begin{document}

\begin{center}
{\bf
SEARCH FOR $Z$-SCALING VIOLATION\\[0.35cm]
IN $p-p$ AND $p-A$ COLLISIONS
 AT HIGH ENERGIES}


\vskip 5mm

 D. Toivonen $^{\S}$ and  M. Tokarev $^{\natural}$

{\small
\vskip 0.3cm
{\it Laboratory of High Energies,\\
Joint Institute for Nuclear Research,\\
141980, Dubna, Moscow region, Russia}

$^{\S}$ {\it  E-mail: {toivonen@sunhe.jinr.ru}}

$^{\natural}$ {\it  E-mail: {tokarev@sunhe.jinr.ru}}
}
\end{center}

\vskip  0.5cm


\begin{center}
\begin{minipage}{150mm}
\centerline{\bf Abstract}
New analysis of experimental data on hadron
$(\pi^{\pm}, K^{\pm}, \bar p) $ production at high-$p_T$
in $p-p$ and  $p-D$ collisions in $z$-presentation
is performed. Data on inclusive cross sections of
cumulative particles produced in backward hemisphere
in $p-A$ collisions  are analyzed as well. The scaling function for
nuclear targets $(Li, Be, C, Al, Cu$ and $ Ta)$ are constructed
and compared with high-$p_T$ data $z$-presentation.
The hypothesis on $z$-scaling violation as a signature
of new physics phenomena is discussed.
\end{minipage}
\end{center}


  {\section{Introduction}}

  The numerous results obtained from data analysis for high-$p_T$
  particle production in $p-p$ and $p-A$ collisions at high energy show
  that $z$-scaling
\cite{Z96}-\cite{Z00}
 \footnote{See contributions
 presented by I.Zborovsk\'{y}, G.\v{S}koro and M.Tokarev}
 reflect general  properties of particle formation.
 Data $z$-presentation can be obtained
  using the experimental observables, the momentum and
  the inclusive cross section $Ed^3\sigma/dq^3$ of produced particle
  at given energy $\sqrt s$ and the multiplicity density
  of charged particles  $dN/d{\eta}|_{\eta=0}=\rho(s)$.

  As argued in  \cite{Z96,Z99,Dedovich,Rog1,Z01}  the scaling function
  $\psi(z)$ describes the  probability density to form a particle
  with formation length $z$. The scaling variable $z$ reveals the property
  of fractal measure. It means that the formation length increases with
  the scale resolution.

  Therefore the violation of $z$-scaling  is of interest to search for
  new phonomena in particle and nuclear physics.

  In the paper experimental data \cite{Jaffe}
  for charged hadrons ($\pi^{\pm}, K^{\pm}, \bar p$)
  produced in $p-p$ and $p-D$ at
  $p_{lab}=400$ and $800~GeV/c $ obtained at Fermilab are analyzed
  and compared with results obtained at  $p_{lab}=70, 200, 300$
  and $400~GeV/c $ before.

  Data sets \cite{Nikif} for cumulative particle
  production in $p-A$ collisions
  are analyzed in the $z$-presentation as well.
  Assuming  that the shape of the scaling curve $\psi(z)$
  is the same as  for high-$p_T$ data
  \cite{Jaffe,Cronin,Protvino}
  the angular  dependence of multiplicity particle density
  for backward hemisphere  particles production \cite{Nikif} is found.

  The obtained results allowed us to compare different data sets
   and point out the kinematical regions interested for search for
  new physics phenomena.


\vskip 0.5cm
{\section{ Hadron production in $p-p$ and $p-D$ at high-$p_T$}}

   Experimental data sets  \cite{Cronin,Protvino} and \cite{Jaffe}
   of inclusive cross sections for $\pi^{\pm}, K^{\pm}, \bar p$ hadrons
   produced  in $p-p$ and $p-D$ collisions at high transverse
   momentum $p_T$  are analyzed.
   The experimental data \cite{Protvino} includes the cross
   sections for $p-p$ and  $p-D$ collisions for
   $\pi^{\pm}, K^{\pm}, p^{\pm}$ hadrons produced at
   $70~GeV/c$  over the  range $1.5<p_T<4.0~GeV/c$.
   The measurements were made at an angle corresponding
   to approximately  $90^0$ in the nucleon-nucleon center-of-mass frame.
   Data  \cite{Jaffe} used in our
   new analysis corresponds to cross sections for
   $\pi^{\pm}, K^{\pm}, p^{\pm}$ hadron
   production in $p-p$  and $p-D$ collisions at $p_{lab} = 400$ and
   $ 800~GeV/c$.
   The produced particles  were registered over the
   transverse momentum  range of $p_{T} = 4.0-10.0~GeV/c$
   and at $\theta_{cm}^{N N}\simeq 90^0$.


  \vskip 0.5cm
{\subsection{$p-p$ collisions}}

 In this section  we study the properties of $z$-scaling
 for hadrons produced in $p-p$  collisions.
We verify the hypothesis of energy scaling for data
$z$-presentation for  hadron  production
in $p-p$ collisions using the available experimental data
 \cite{Cronin,Protvino}  and  \cite{Jaffe}.
The energy scaling for data  $z$-presentation means that the
shape of the scaling curve is  independent of the collision energy
$\sqrt s$.

 Figures 1(a) and 2(a)  show the dependence of the cross section
 $Ed^3\sigma/dq^3$  of $\pi^{+}$ and $K^{-}$ produced in $p-p$
 on transverse momentum $p_{T}$ at $p_{lab} = 70, 200, 300, 400$
 and $800~GeV/c$ and the angle $\theta_{cm}$ near $90^0$.
 Note that the data cover the wide transverse momentum
 range, $p_{T}=1-10~GeV/c$.

 The hadron spectra have a power behavior and
 demonstrate the strong energy dependence
 increasing with the transverse momentum.

 Figures 1(b) and 2(b) show $z$-presentation of the same data sets.
 Taking into account the experimental errors we can conclude that
 the scaling function $\psi(z)$ demonstrates an energy
 independence over a wide energy and transverse momentum
 range at $\theta_{cm} \simeq 90^0$.

 We found that new data \cite{Jaffe} included in the
 analysis give no indications on $z$-scaling violation.
 The result is the new confirmation of energy independence of data
 $z$-presentation.

\vskip 0.5cm
{\subsection{$p-D$ collisions}}

The study of the energy independence of data $z$-presentation
for $p-D$ collisions is especially interest. It is connected with
the possibility to investigate the influence of nuclear
matter on particle formation. We assume that in the high-$p_T$
range the effects can give direct information
on the state of nuclear medium which determines the
properties of particle formation.
Moreover as shown in \cite{Z01}  due to
A-dependence of data $z$-presentation
the scaling functions
for different nuclei from $D$ to $Pb$ are coincide each other
if the scaling transformation
$ z \rightarrow \alpha (A) \cdot z,$  \
$ \psi \rightarrow \alpha^{-1} (A) \cdot \psi $
depending on the single parameter, the atomic weight $A$ is used.
It means that the lightest nucleus, deuterium, can be a good target
to search for a signature of nuclear phase transition at high-$p_T$.
Let us remind that the quantitative measure of $z$-scaling
violation and consequently one of the possible signatures
of new physics phenomena
is the change of the anomalous fractal dimension $\delta$.

Figures 3(a) and 4(a) show $p_T$-presentation of inclusive cross section
$Ed^3\sigma/dq^3$ of $\pi^{+}$ and $K^{-}$ produced in $p-D$
at $p_{lab} = 70$ and $400~GeV/c$
and the angle $\theta_{cm}^{NN}\simeq 90^0$.
Note that the data \cite{Jaffe} are in a good agreement with
the data obtained by Cronin group \cite{Cronin} at $p_{lab}=400~GeV/c$.

As  seen  from  Figures  3(b) and 4(b)  $z$-presentation
for  all  data  sets \cite{Cronin,Protvino} and \cite{Jaffe}
demonstrate the energy independence.
It means  that $A$-dependence of $z$-presentation
is correctly described  by the function $\alpha(A)=0.9A^{0.15}$
established in \cite{Z01}.

Thus we can conclude that
data \cite{Jaffe} for $p-D$ collisions included in the
new analysis do not give any indications on $z$-scaling
violation up $z\simeq 20$.

\vskip 0.5cm
{\subsection{$z-p_T$ plot}}

To determine the kinematical region preferable for searching for the
scaling violation $z-p_T$ plot is suggested  to use.

Figure 5  shows  the  dependence of $z$ as a function of transverse
momentum $p_T$  for  $p-p$ (a)  and $p-D$ (b)  collisions at
$p_{lab} = 70, 200, 400$ and $800~GeV/c$ and
$\theta_{cm} \simeq 90^0$.
The results of present analysis allow us to conclude that
the systematic experimental study of high-$p_T$ particle spectra
for $p-p$ and $p-D$ collisions at $z > 10 $ is necessary to determine
the asymptotic behavior of the scaling function $\psi(z)$.
The kinematical region $ z > 10$ is more preferable to search for
$z$-scaling violation.

\vskip 0.5cm
{\section{Cumulative particle production in $p-A$}}

Cumulative particles are called particles produced in the
kinematical region forbidden for free nucleon-nucleon interaction
\cite{Stavinsky,Zolin}-\cite{ABaldin}.
Such particles can be only produced in the processes with participation
of nuclei (in hadron-nucleus, nucleus-nucleus and lepton-nucleus collisions).

 Inclusive cross sections $Ed^3\sigma/dq^3$  for
 $\pi^{\pm}, K^{\pm}, p^{\pm}$ hadron
 production in backward hemisphere in  $p-A$
 collisions at $p_{lab} = 400~GeV/c$ and
 at the angle  $\theta_{lab}$ of $ 70^0,
 90^0, 118^0$ and $160^0$ are presented in \cite{Nikif}.
 The measurements were performed over the  momentum
 range $0.2<p<1.25~GeV/c$. Nuclear targets, $Li, Be, C, Al,Cu$
 and $Ta$ were used.
 The data cover in particular the kinematical range
 forbidden for particle production in nucleon-nucleon collisions.

In the paper we restrict ourselves of the analysis of data
\cite{Nikif} obtained at Batavia. More complete analysis of
cumalative data sets will be presented elsewhere.

Figures 6(a) and 7(a) present the inclusive cross sections for
$\pi^{+}$ mesons  produced in backward hemisphere in $p-Be$ and $p-Ta$
collisions at $p_{lab} = 400~GeV/c$ and
 the angle  $\theta_{lab}$ of $ 70^0,
90^0, 118^0$ and $160^0$.
As seen from Figures 6(a) and 7(a)
the strong  dependence of cross sections
as $\theta_{lab}$ changes from $70^0$ to $160^0$
is observed for all nuclear targets, $Li, Be, C, Al, Cu, Ta$.

We assume that the shape of the scaling curve will be the same
as for data points corresponding to the high-$p_T$ region.
At present there are not experimental data on
the angular dependence of  $\rho(s,\eta, A)$ for particles
produced in backward hemisphere in order to construct directly
the scaling function.
Therefore we study the possibility to describe the shape of the $\psi(z)$
found from the analysis of high-$p_T$ data sets
\cite{Protvino,Cronin,Jaffe}   using data points \cite{Nikif}.
The function  $\rho(s,\eta, A)$ is parameterized in the form
$\rho(s,\eta, A)= \rho(s,A)|_{\eta=0} \cdot \chi(\theta, A)$, where
the angular dependence is described by  $\chi(\theta, A)$.
The latter is shown in Figure 8(a).
As seen from Figure the ratio
$\chi_{Ta}/\chi_{Li}$  decreases from 3.5 to 1.5 as
the angle $\theta_{lab}$ increases from $70^0$ to $160^0$.
The $A$-dependence of the ratio demonstrates the saturation
reached for backward particle production.

Figures 6(b) and 7(b)  show the $z$-presentation of data
\cite{Nikif}. One can see that the curve found
for nuclei  $ Be$ and $ Ta$ is in a good
agreement  with  high-$p_T$  data $z$-presentation for D.
Note that the angular dependence  of
$\chi(\theta, A)$  shown in Figure  8 (a)
describes  both non- and cumulative data points
\cite{Nikif}.
Thus we found that there is possibility to combine
the scaling functions corresponding to
the cumulative and  high-$p_T$ data.
As seen from Figures 6(b) and 7(b)
all points \cite{Nikif} are out of the asymptotic region.
Therefore it is of interest to determine the kinematical region
for backward pion production where $z_{cum} >  z_{hard} \simeq 20$.

\vskip 0.5cm
{\subsection { $z-p$ plot}}

 The dependence of $z$ on momentum $p$ ( $z-p$  plot)
 as a function of an atomic weight $(A)$, an angle of produced
 particle $(\theta)$ and a collision energy $\sqrt s $
 can be used  to select the domain where
 the scaling can be violated and  new physical
 phenomena can be found.

  Figure  8(b)  shows a $z-p$ plot for  $p-Ta$
  collisions at $p_{lab} = 70, 200, 400 $ and $800~GeV/c$ and
  $\theta_{lab}=160^0$.
  The value $z=20$
  for $p-Ta$ collisions  corresponds to the values of the
  momentum $p$ of 1.9, 2.1, 2.2 and $2.25~GeV/c$, at
  $p_{lab} = 70, 200, 400$ and $800~GeV/c$, respectively.

\vskip  0.5cm
{\section{Conclusions}}

To  search for $z$-scaling violation
the analysis of the scaling features of
 hadrons produced in $p-p$ and $p-A$
collisions at high energies in
terms of $z$-presentation are performed.
The experimental high-$p_T$  \cite{Jaffe,Cronin,Protvino}
and cumulative  \cite{Nikif} data  sets on the inclusive
cross sections are used in the analysis.


  Data $z$-presentation for high-$p_T$ and cumulative data
  is constructed. It is expressed via the
  experimental observables, momenta and masses of
  colliding and produced particles,
  the invariant inclusive cross section
  $Ed^3\sigma/dq^3$  and the multiplicity  density of charged
  particles.

 A new confirmation of $z$-scaling for particle production
 in high-$p_T$ range is obtained.
  It is shown that available experimental high-$p_T$ data
  \cite{Protvino,Cronin,Jaffe} on hadron production in $p-p$ and $p-D$
  collisions give no indications on $z$-scaling violation.
  The shape of the scaling curve does not change with  the collision
  energy $\sqrt s$.

  The found angular dependence of particle multiplicity density
  for backward  pion  production  in $p-A$  reproduces the
  shape of the scaling curve obtained from high-$p_T$  data.



 New data \cite{Jaffe} included in the present analysis
 point out the asymptotic regime of the scaling function,
 $\psi(z)\sim  z^{-\beta}$,  at $z>4$.

The kinematical ranges for high-$p_T$ and backward hemisphere
particle production preferable  for search for $z$-scaling violation
in $p-p$ and $p-A$ collisions are found using the $z-p_T$ and $z-p$ plots.




Analysis of data on cumulative particle production has
shown a possibility to use the class of events to search for
$z$-scaling violation. It is assumed  that the properties
of nuclear matter in the
cumulative range should drastically change the mechanism of particle
formation due to multiple interactions of elementary constituents.
Therefore the systematic study of particle spectra at high-$z$
and multiplicity particle density in the backward hemisphere
is of interest.

%

%

\vskip 1.5cm
{\large \bf Acknowledgments}

\vskip 0.5cm
The authors would like to thank I.Zborovsky and O.Rogachevsky
for useful discussions of the present work.

%
%

{\small

\newpage

\begin{figure}[htb]

\hspace*{-4cm}
\centerline{\epsfig{file=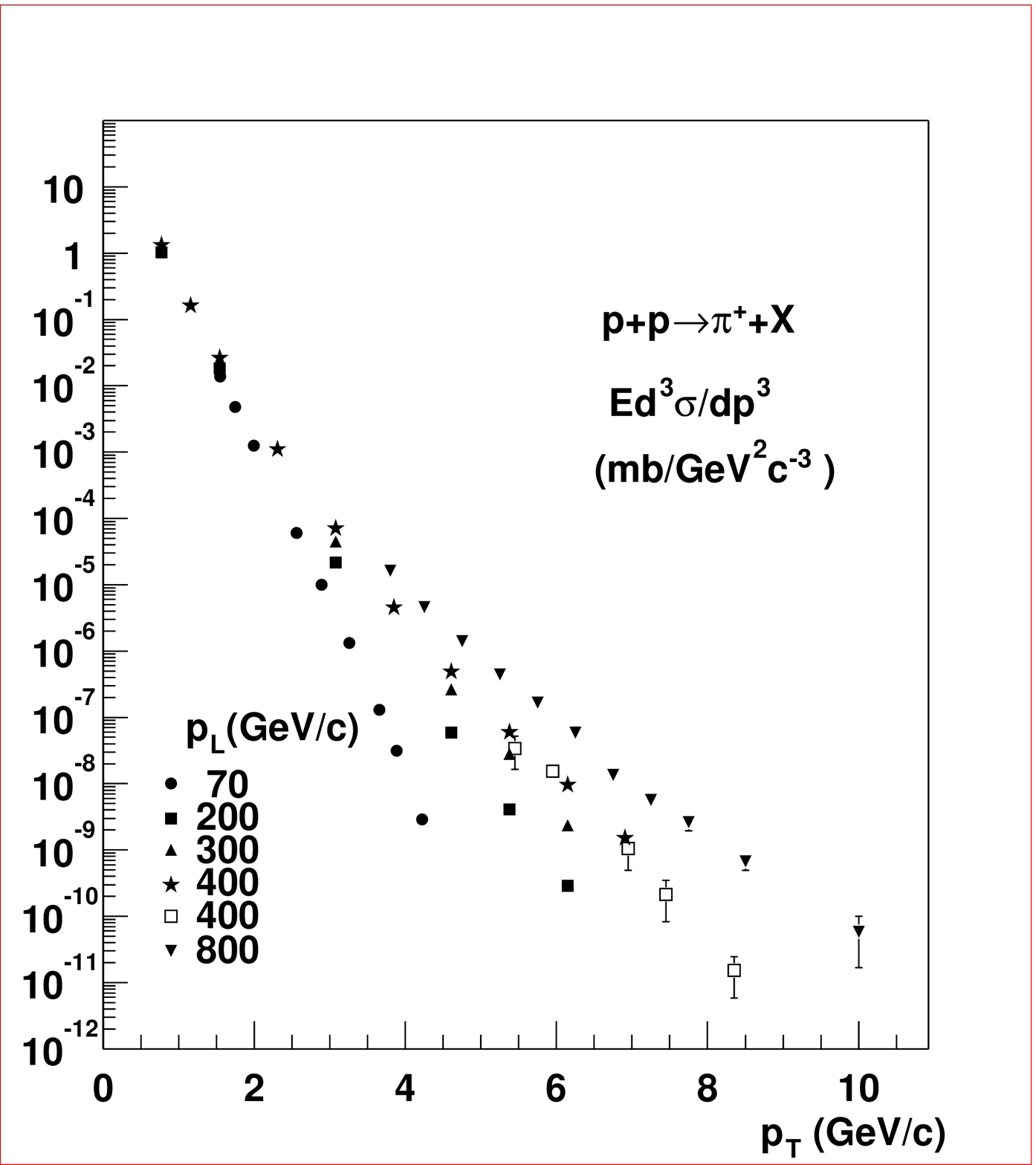,width=7.0cm}}
\vskip -7.8cm
\hspace*{4cm}
\centerline{\epsfig{file=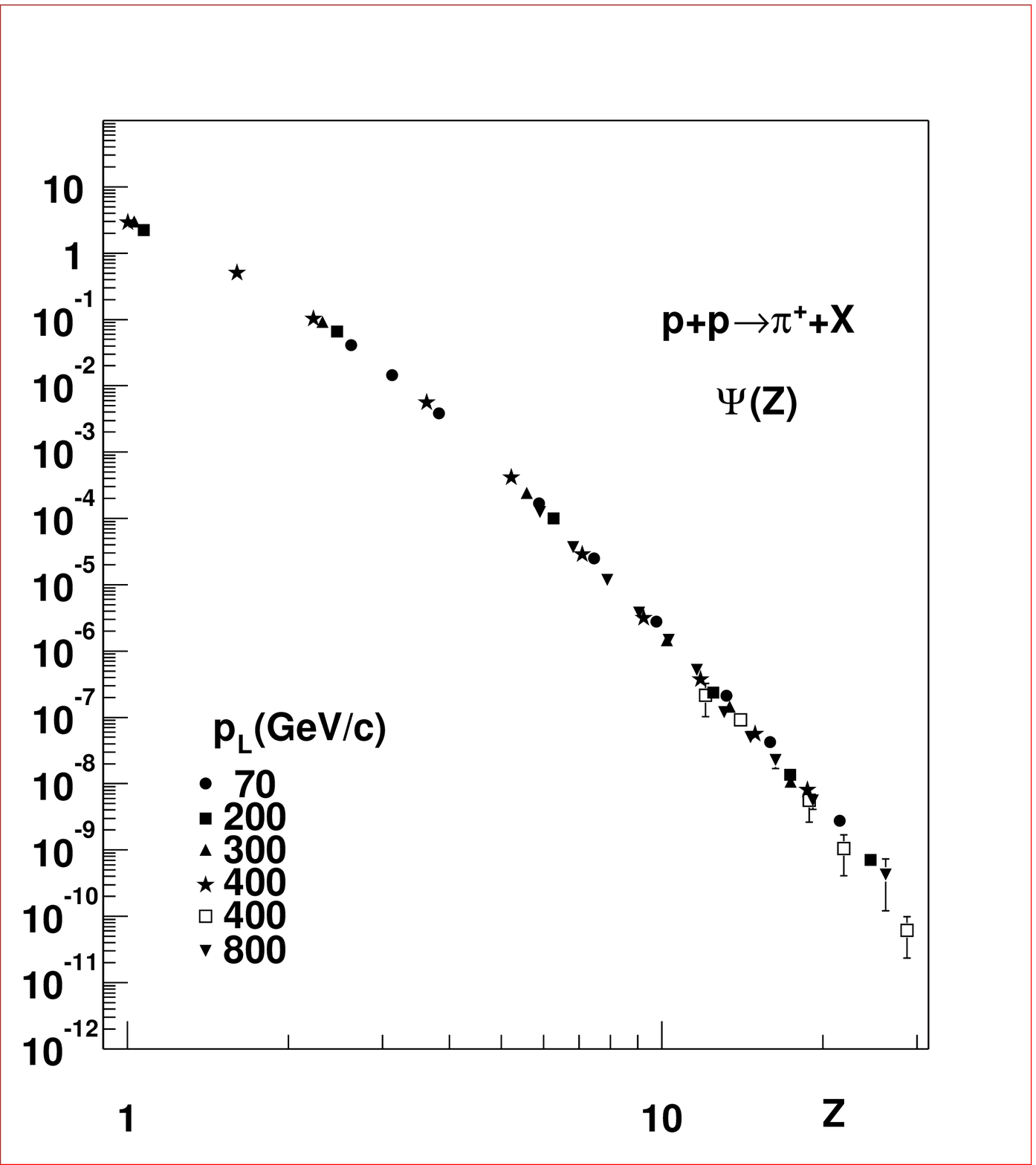,width=7.cm}}
\vskip 0.5cm
\hspace*{4.cm} a) \hspace*{8.cm} b)\\[0.0cm]
{{\bf Figure 1.}
 (a) Dependence of  the
 inclusive cross section of $\pi^+$-meson  production in  $p-p$ collisions
 on transverse momentum $p_{T}$ at $p_{lab} = 70, 200,300,400$ and
 $800~GeV/c$
 and $\theta_{cm} \simeq 90^{0}$.
 Experimental data are taken from
 \cite{Jaffe,Cronin,Protvino}.
 (b) The corresponding scaling function $\psi(z)$. }
\end{figure}


\begin{figure}[htb]

\hspace*{-4cm}
\centerline{\epsfig{file=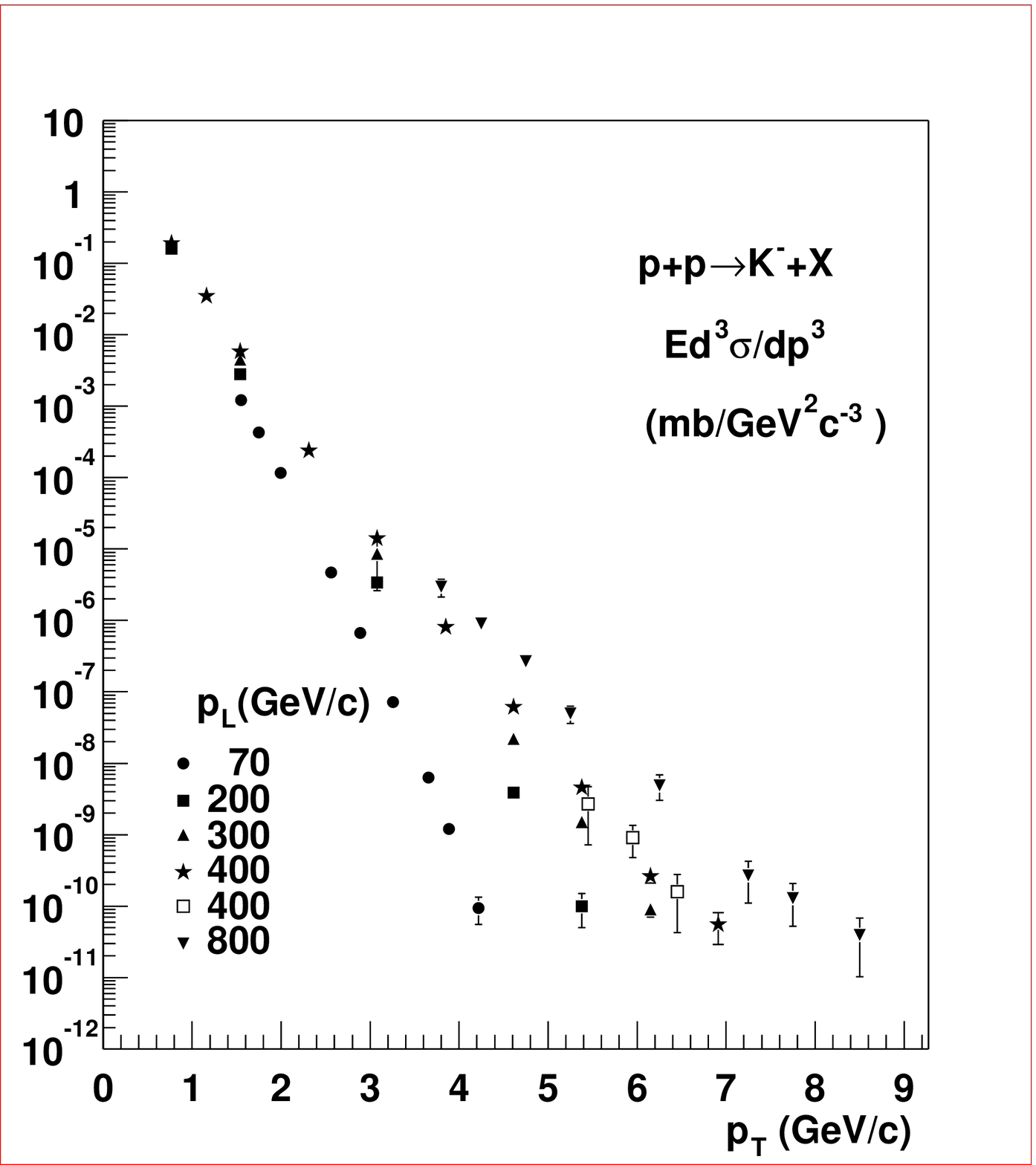,width=7.0cm}}
\vskip -7.8cm
\hspace*{4cm}
\centerline{\epsfig{file=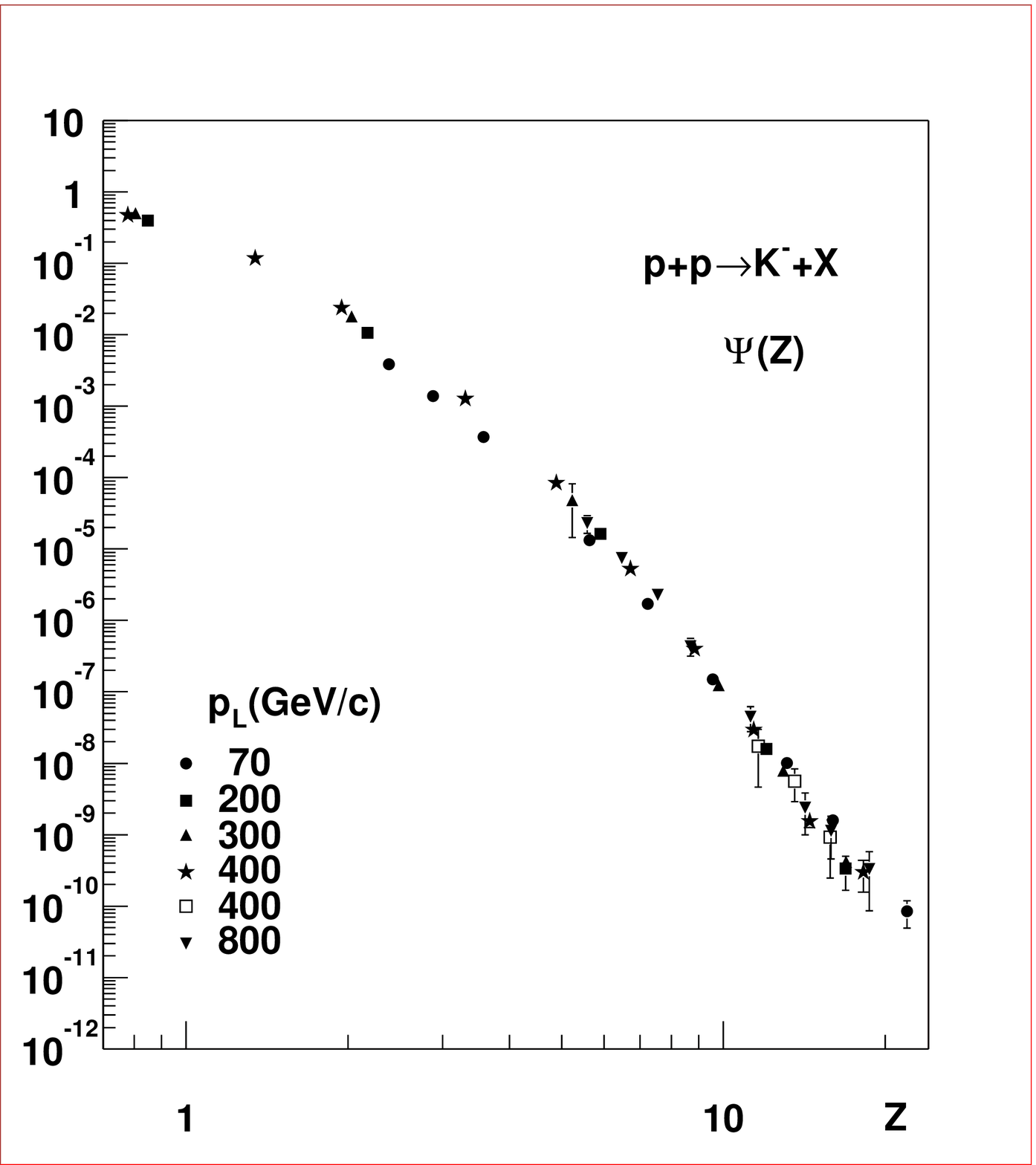,width=7.cm}}
\vskip 0.5cm
\hspace*{4.cm} a) \hspace*{8.cm} b)\\[0.0cm]
{{\bf Figure 2.}
 (a) Dependence of  the
 inclusive cross section of $K^-$-meson  production
 in  $p-p$ collisions
 on transverse momentum $p_{T}$ at $p_{lab} = 70, 200,300,400$ and
 $800~GeV/c$
 and $\theta_{cm} \simeq 90^{0}$.
 Experimental data are taken from
 \cite{Jaffe,Cronin,Protvino}.
 (b) The corresponding scaling function $\psi(z)$. }
\end{figure}

\newpage

\begin{figure}[htb]

\hspace*{-4cm}
\centerline{\epsfig{file=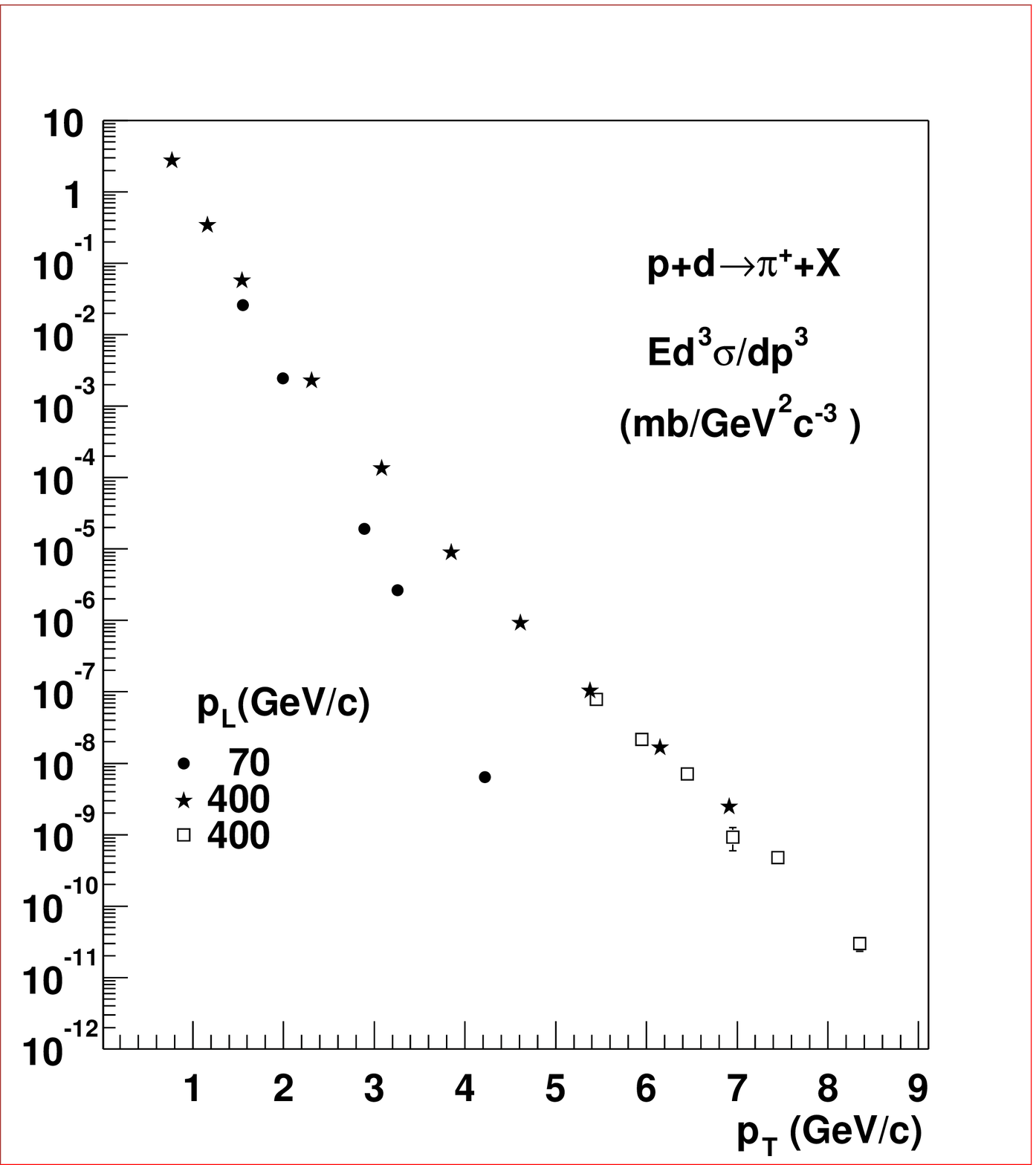,width=7.0cm}}
\vskip -7.8cm
\hspace*{4cm}
\centerline{\epsfig{file=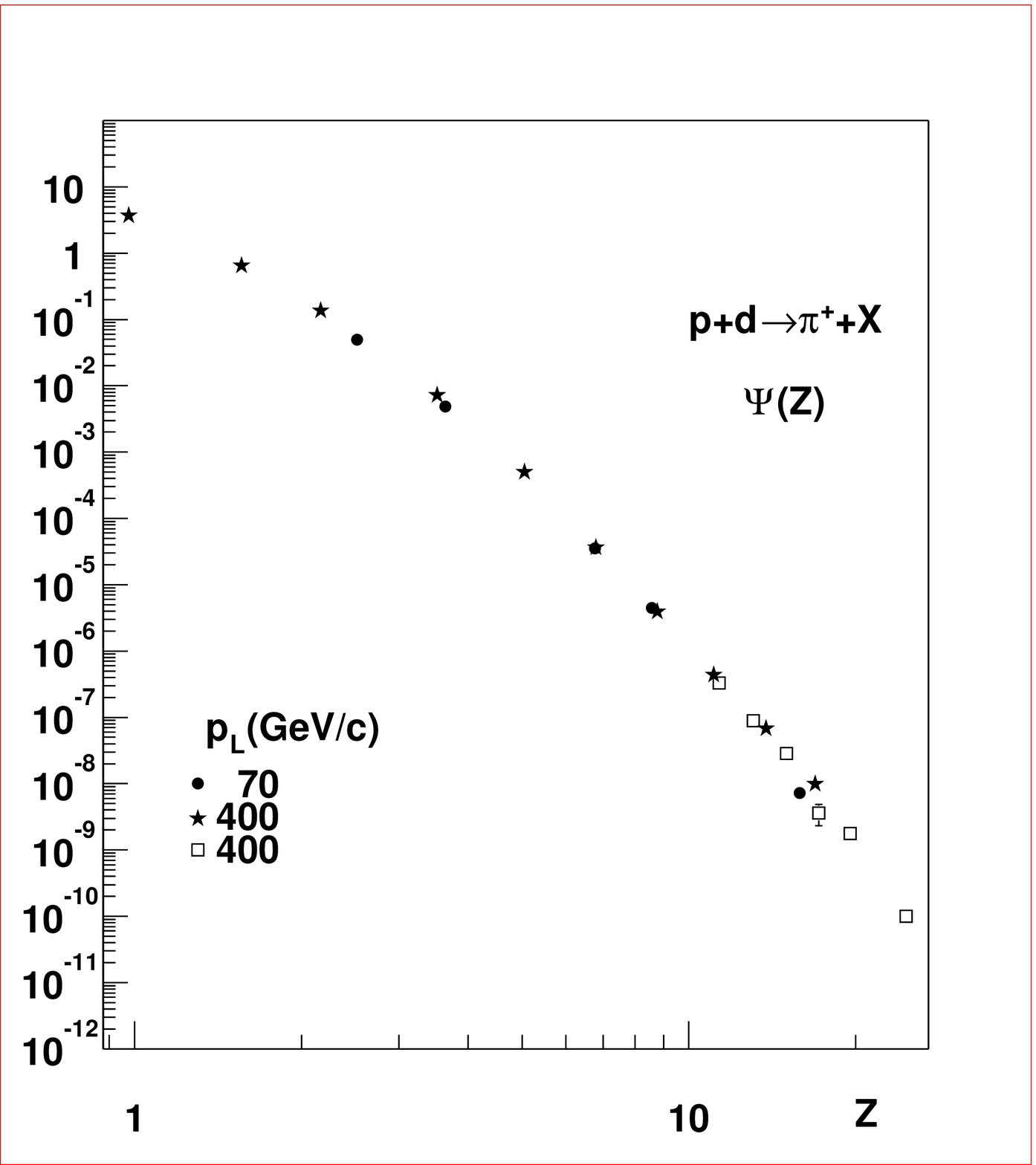,width=7.cm}}
\vskip 0.5cm
\hspace*{4.cm} a) \hspace*{8.cm} b)\\[0.0cm]
{{\bf Figure 3.}
 (a) Dependence of  the
 inclusive cross section of $\pi^+$-meson  production
 in  $p-D$ collisions
 on transverse momentum $p_{T}$ at $p_{lab} = 70$ and
 $400~GeV/c$
 and $\theta_{cm} \simeq 90^{0}$.
 Experimental data are taken from
 \cite{Jaffe,Cronin,Protvino}.
 (b) The corresponding scaling function $\psi(z)$. }
\end{figure}


\begin{figure}[htb]

\hspace*{-4cm}
\centerline{\epsfig{file=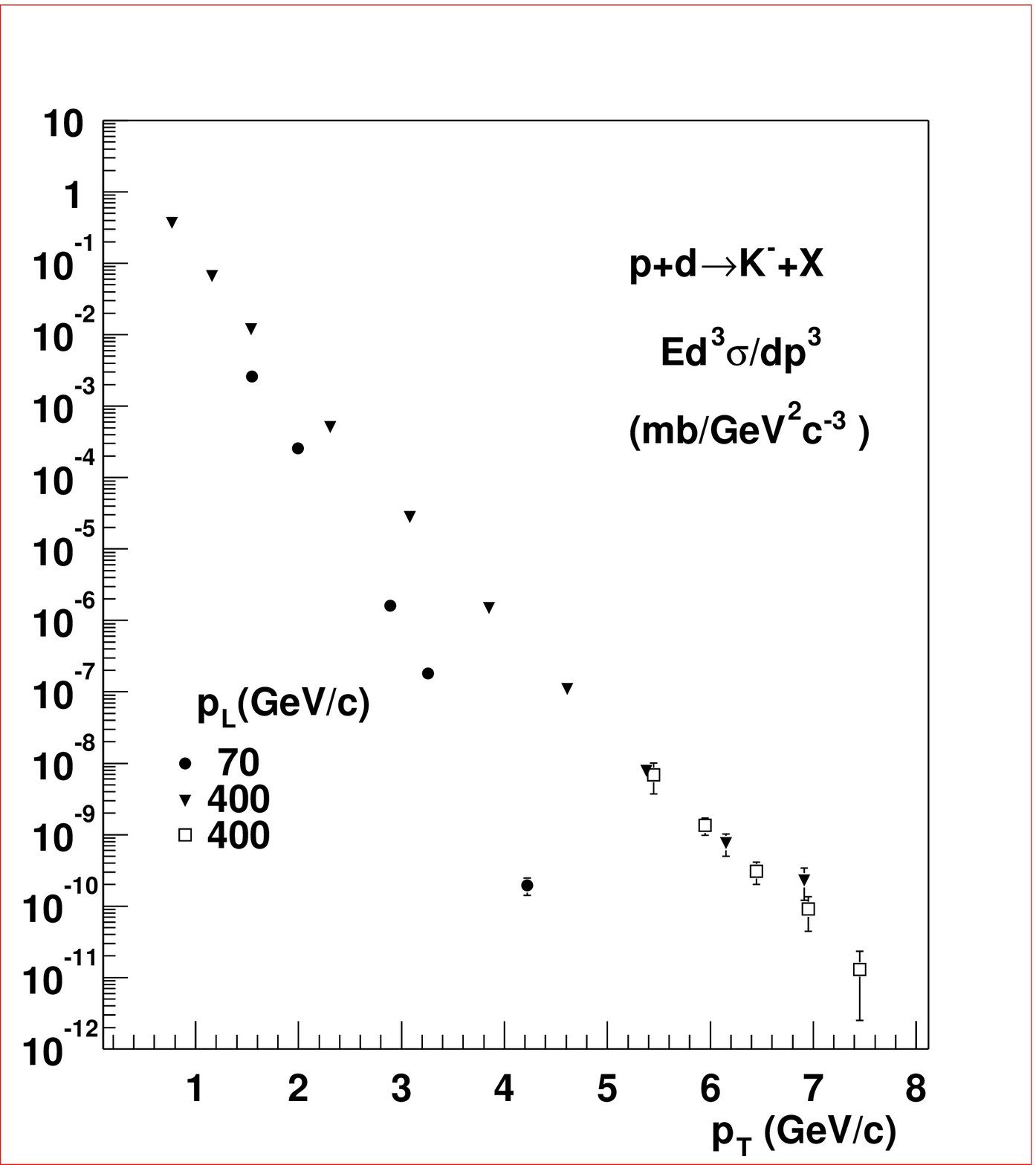,width=7.0cm}}
\vskip -7.8cm
\hspace*{4cm}
\centerline{\epsfig{file=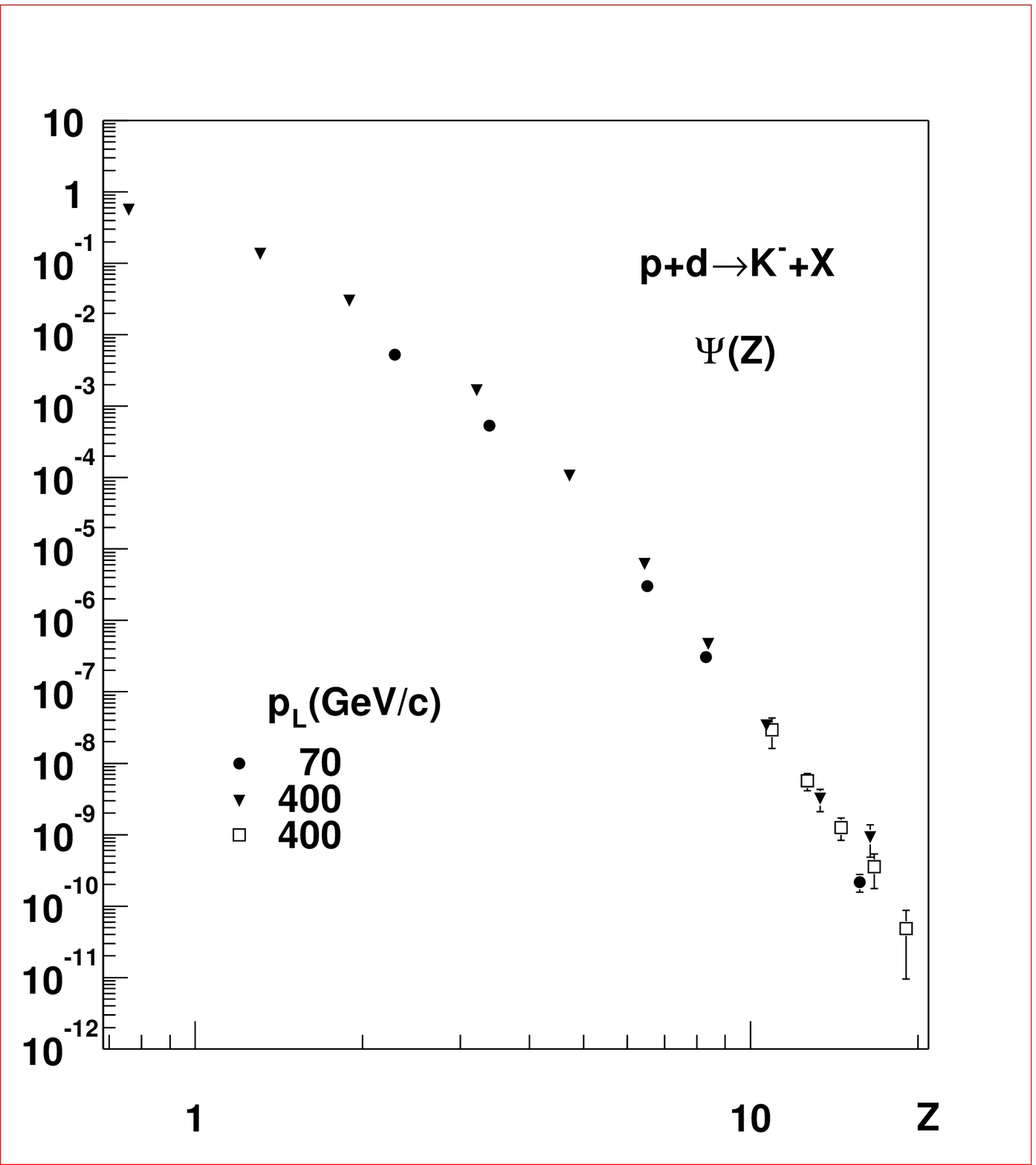,width=7.0cm}}
\vskip 0.5cm
\hspace*{4.cm} a) \hspace*{8.cm} b)\\[0.0cm]
{{\bf Figure 4.}
 (a) Dependence of  the
 inclusive cross section of $K^-$-meson  production
 in  $p-D$ collisions
 on transverse momentum $p_{T}$ at $p_{lab} = 70$ and
 $400~GeV/c$
 and $\theta_{cm} \simeq 90^{0}$.
 Experimental data are taken from
 \cite{Jaffe,Cronin,Protvino}.
 (b) The corresponding scaling function $\psi(z)$. }
\end{figure}

\newpage

\begin{figure}[htb]

\hspace*{-4cm}
\centerline{\epsfig{file=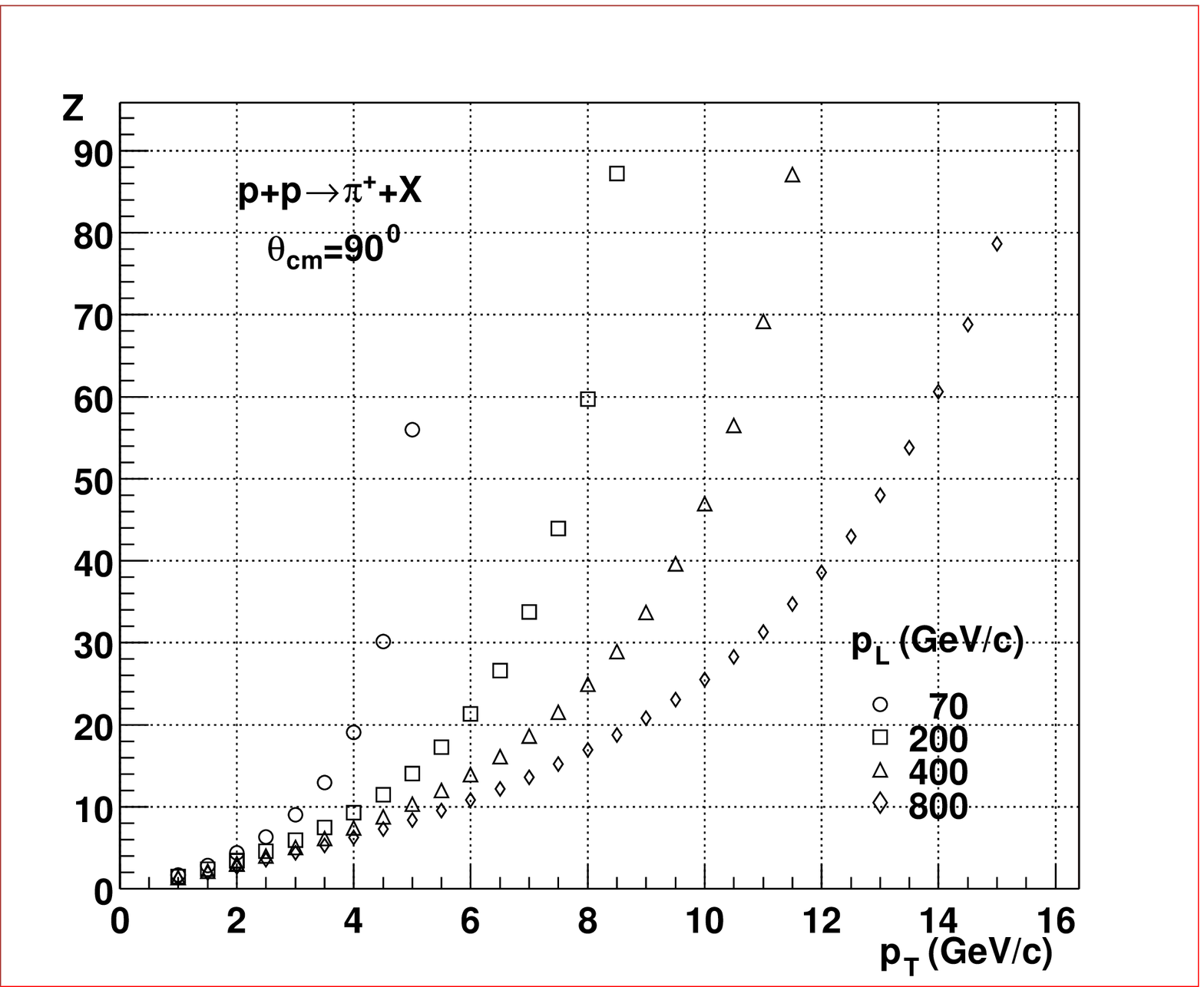,width=7.0cm}}
\vskip -5.8cm
\hspace*{4cm}
\centerline{\epsfig{file=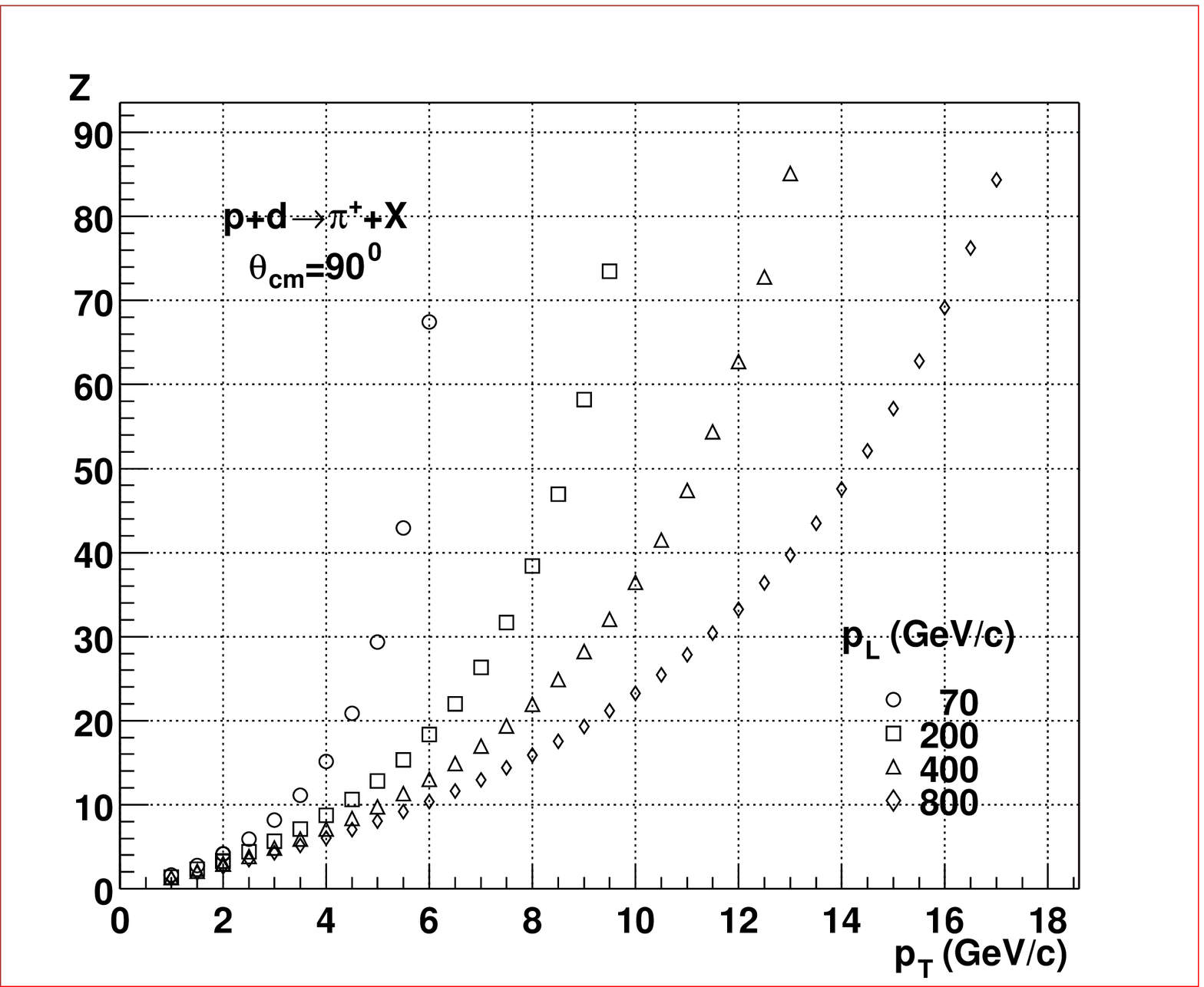,width=7.cm}}
\vskip 0.5cm
\hspace*{4.cm} a) \hspace*{8.cm} b)\\[0.0cm]
{{\bf Figure 5.}
 $z-p_T$ plot for $\pi^+$-meson production in $p-p$ (a) and
$p-D$ (b)
    collisions  at
    $p_{lab} = 70, 200, 400$ and $800~GeV/c$
and $\theta_{cm} \simeq 90^0$. }
\end{figure}


\begin{figure}[htb]

\hspace*{-4cm}
\centerline{\epsfig{file=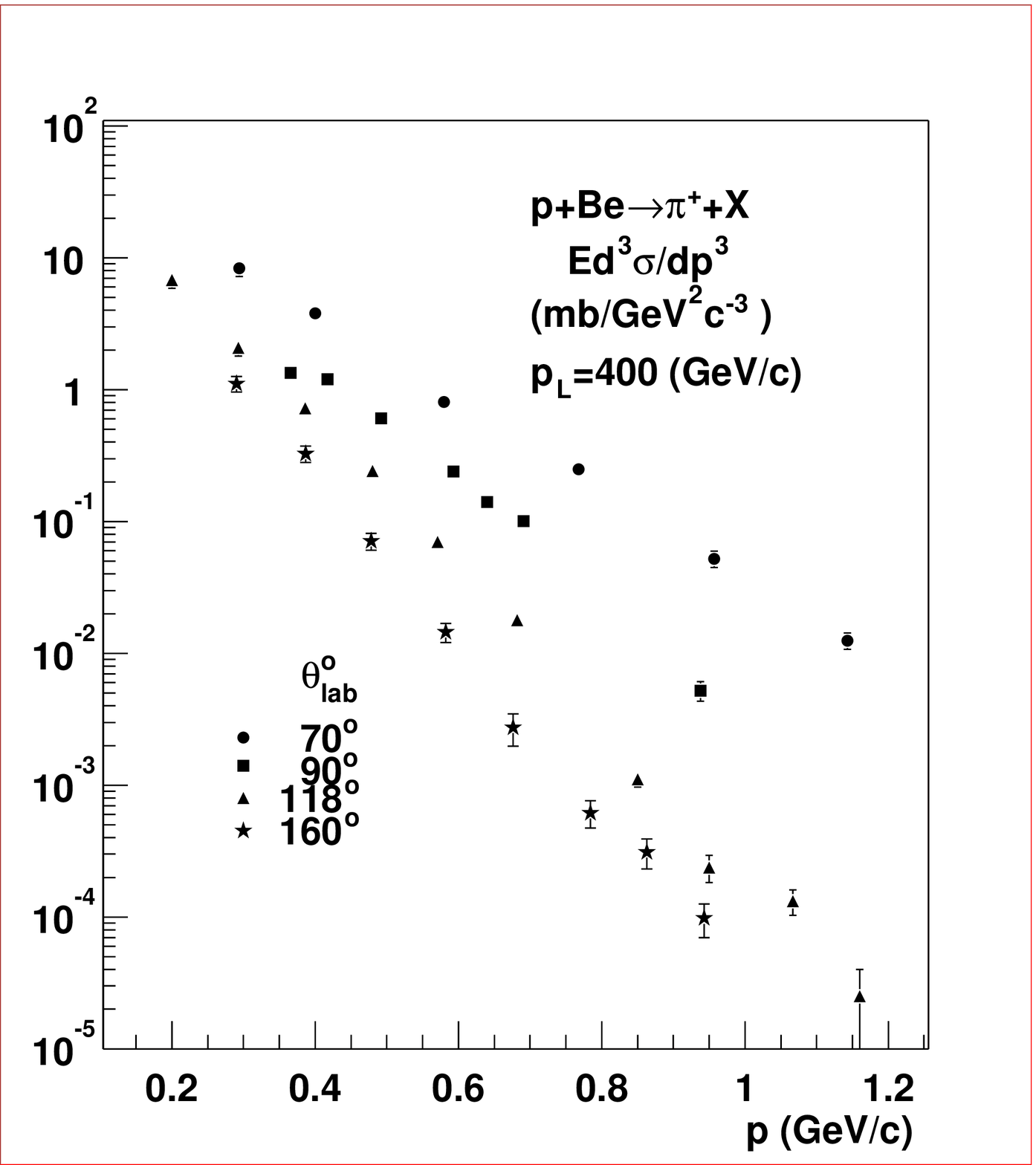,width=7.0cm}}
\vskip -7.8cm
\hspace*{4cm}
\centerline{\epsfig{file=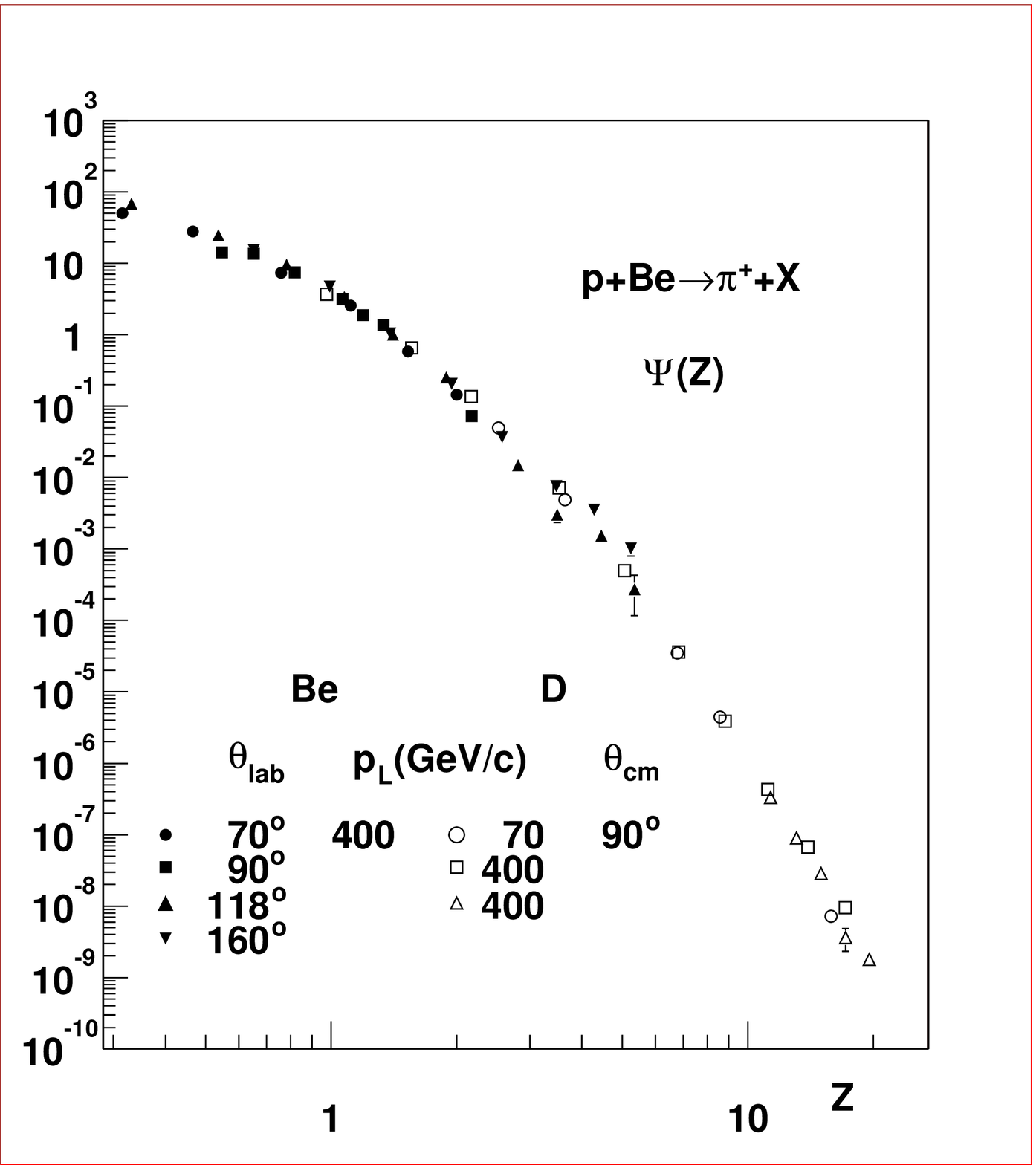,width=7.cm}}
\vskip 0.5cm
\hspace*{4.cm} a) \hspace*{8.cm} b)\\[0.0cm]

{{\bf Figure 6.}
 (a) Dependence of the
 inclusive cross section of $\pi^+$-meson  production
 in  $p-Be$ collisions
 on momentum $p$ at $p_{lab} = 400~GeV/c$
 and $\theta_{lab}=70^0, 90^0, 118^0, 160^0$.
 Experimental data are taken from \cite{Nikif}.
 (b) The corresponding $z$-presentation  of data sets
\cite{Nikif} and
\cite{Jaffe,Cronin,Protvino}.
 }
\end{figure}

\newpage

\begin{figure}[htb]

\hspace*{-4cm}
\centerline{\epsfig{file=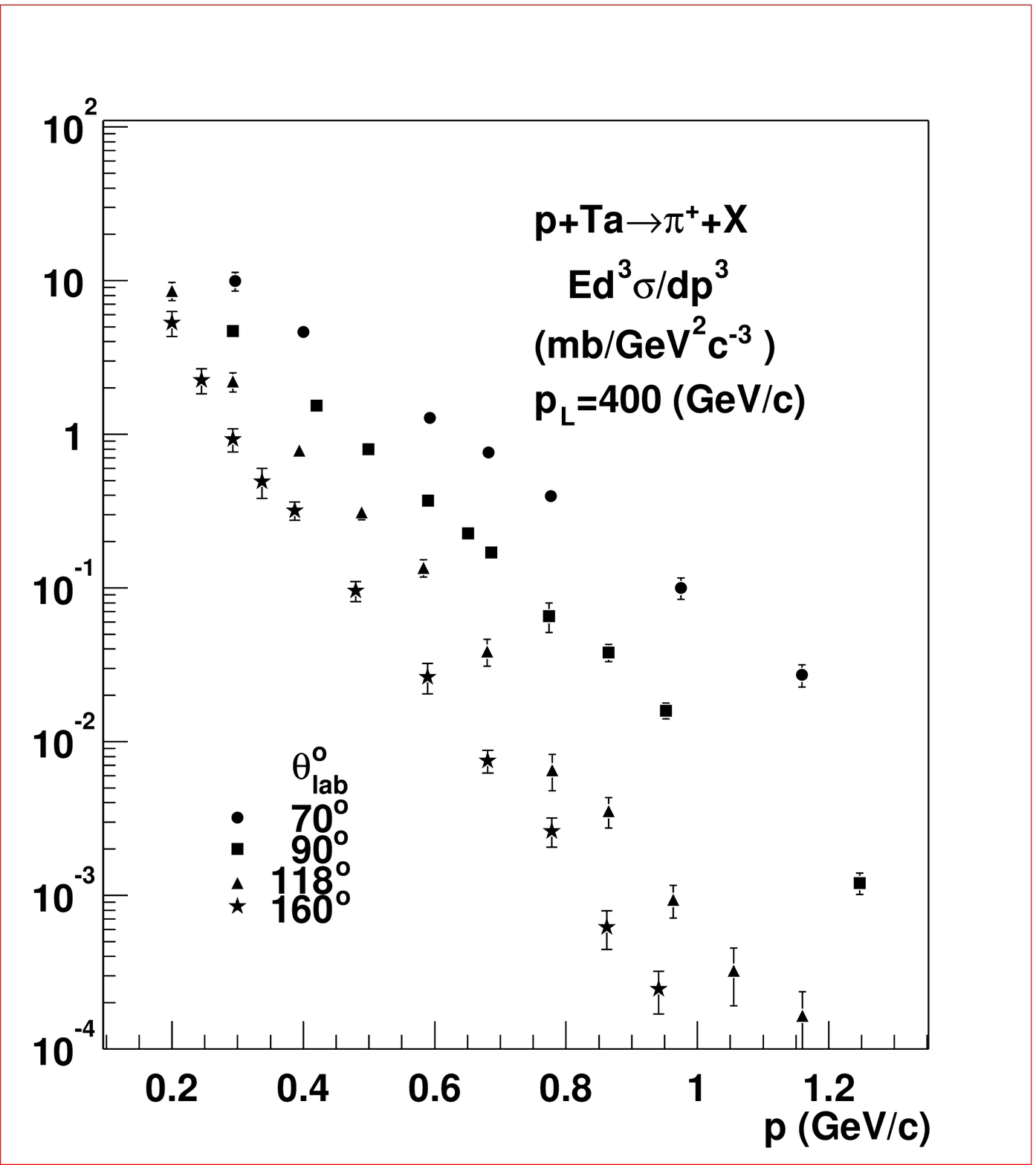,width=7.0cm}}
\vskip -7.8cm
\hspace*{4cm}
\centerline{\epsfig{file=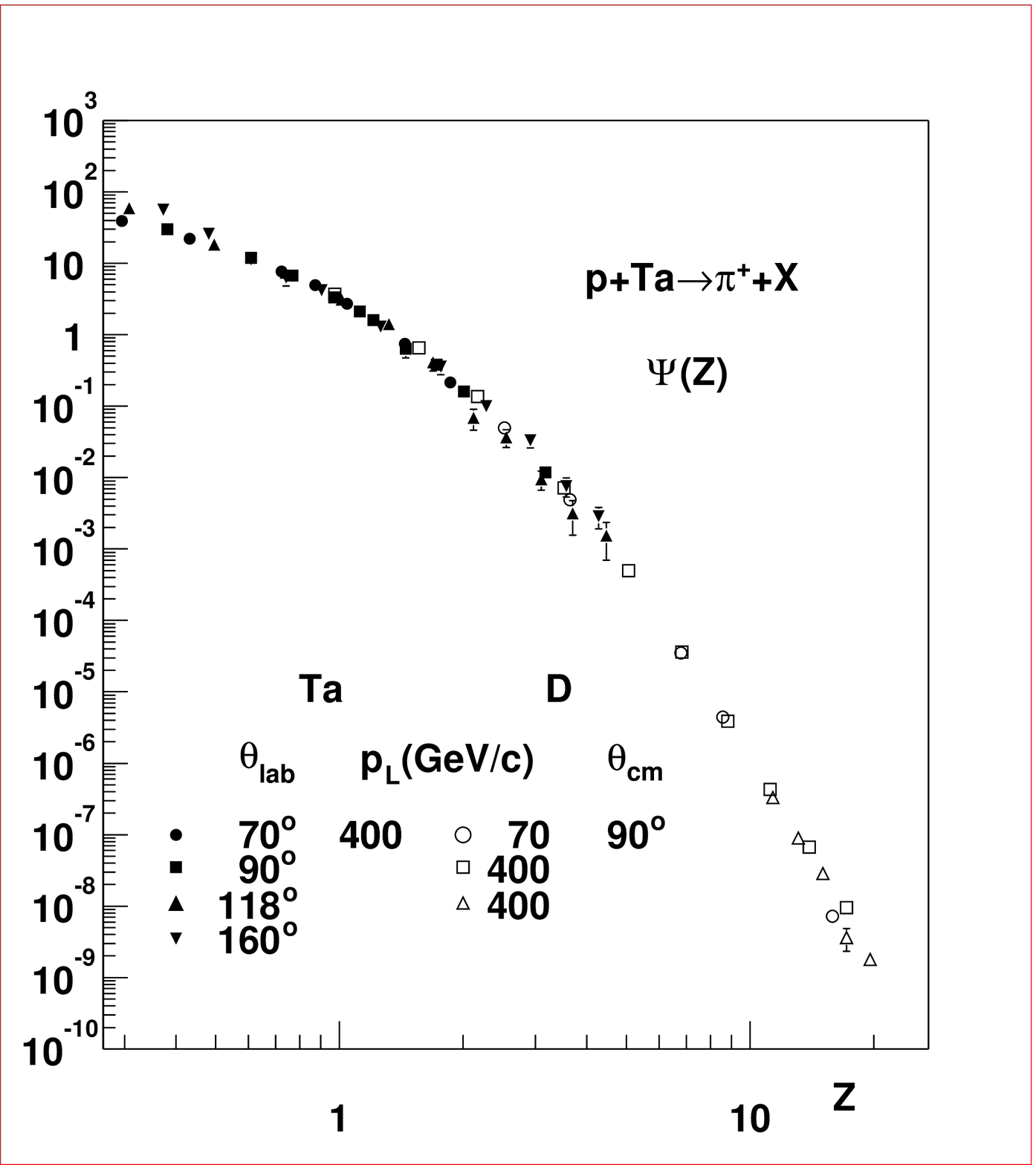,width=7.cm}}
\vskip 0.5cm
\hspace*{4.cm} a) \hspace*{8.cm} b)\\[0.0cm]
{{\bf Figure 7.}
 (a) Dependence of the
 inclusive cross section of $\pi^+$-meson  production
 in  $p-Ta$ collisions
 on momentum $p$ at $p_{lab} = 400~GeV/c$
 and $\theta_{lab}=70^0, 90^0, 118^0, 160^0$.
 Experimental data are taken from \cite{Nikif}.
 (b) The corresponding $z$-presentation  of data sets
\cite{Nikif} and \cite{Jaffe,Cronin,Protvino}.  }
\end{figure}


\begin{figure}[htb]

\hspace*{-4cm}
\centerline{\epsfig{file=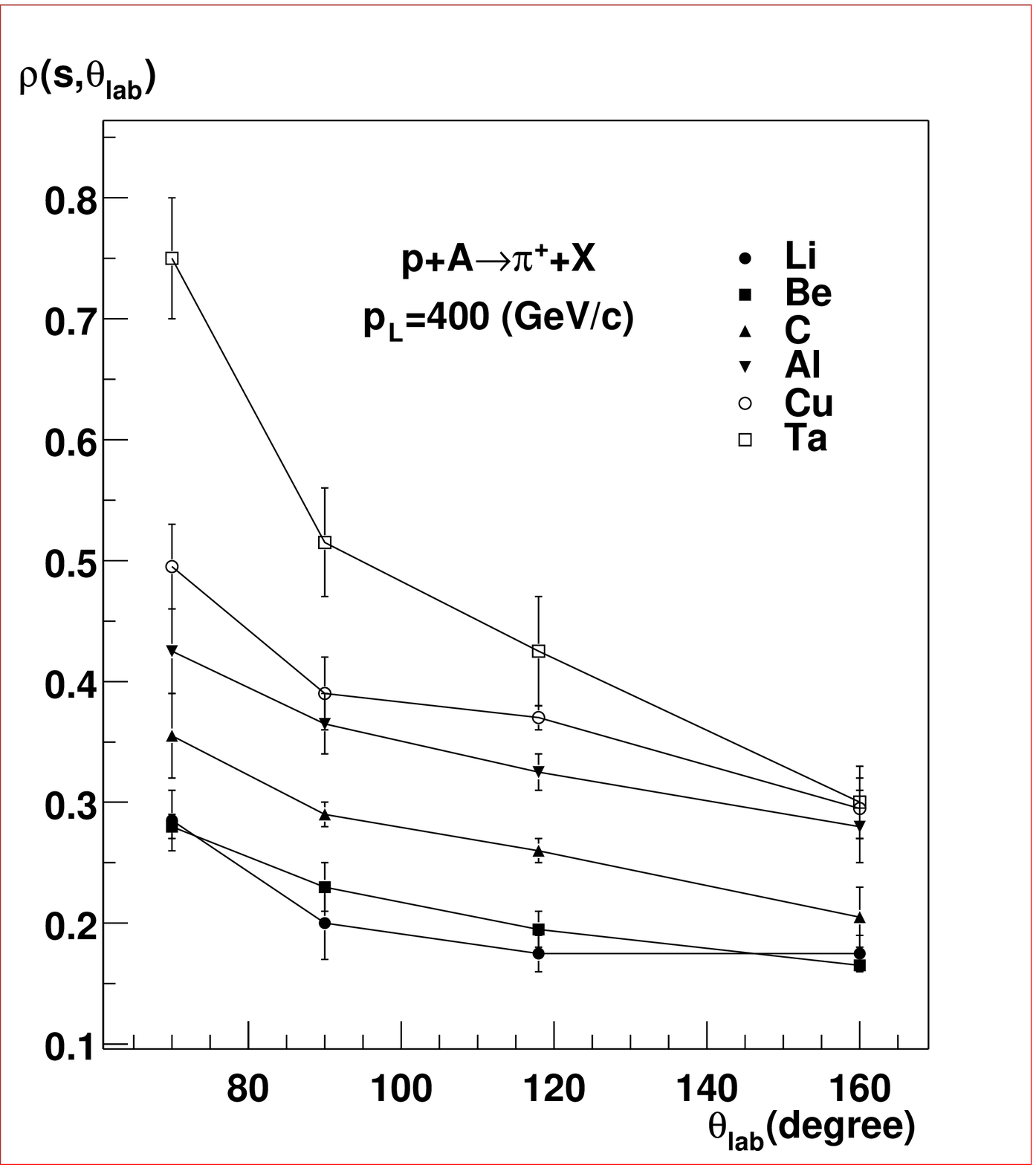,width=7.0cm}}
\vskip -6.8cm
\hspace*{4cm}
\centerline{\epsfig{file=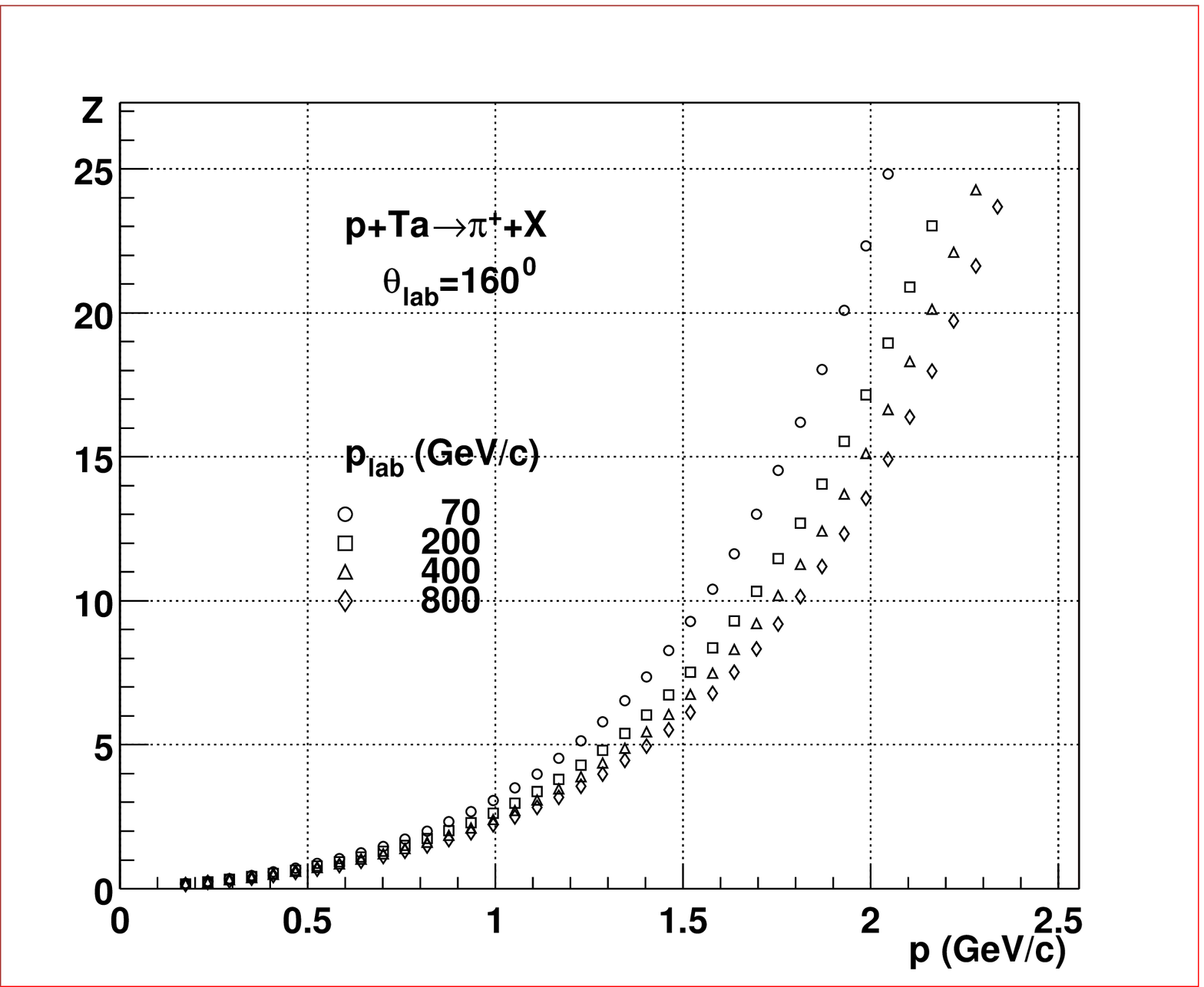,width=7.cm}}
\vskip 1.cm
\hspace*{4.cm} a) \hspace*{8.cm} b)\\[0.0cm]
{{\bf Figure 8.}
 (a)
  Angular dependence $\chi(s, \theta, A)$ of
     particle multiplicity density for particle production
in $p-A$ collisions  at $p_{lab} = 400~GeV/c$
 and $\theta_{lab}=70^0, 90^0, 118^0, 160^0$.
(b)  $z-p$ plot for $\pi^+$-meson production in $p-Ta$
   collisions  at  $p_{lab} = 70, 200, 400$ and $800~GeV/c$
and $\theta_{lab} =160^0$. }
\end{figure}


\begin{thebibliography}{99}





\bibitem{Z96}
I.Zborovsk\'{y}, Yu.Panebratsev, M.Tokarev, G.\v{S}koro, Phys.
Rev. {\bf D54} (1996) 5548.


\bibitem{Z99}
I.Zborovsk\'{y}, M.Tokarev, Yu.Panebratsev, G.\v{S}koro,
Phys. Rev. {\bf C59}, 2227 (1999).






\bibitem{Dedovich}  M.V.Tokarev, T.G.Dedovich,
 Int. J. Mod. Phys. {\bf A15} (2000) 3495.

\bibitem{Rog1}  M.Tokarev, O.Rogachevski, T.Dedovich,
J. Phys. G: Nucl. Part. Phys. {\bf 26} (2000) 1671.






\bibitem{Z01}
 M.Tokarev, I.Zborovsk\'{y}, Yu.Panebratsev, G.\v{S}koro,
 Int. J. Mod. Phys. {\bf A16} (2001) 1281.



\bibitem{Zb}
I.Zborovsk\'{y}, hep-ph/0101018

\bibitem{Z00}
I.Zborovsk\'{y}, M.Tokarev, Yu.Panebratsev, G.\v{S}koro,
  JINR Preprint  E2-2001-41, Dubna, 2001.





 \bibitem{Jaffe} D. Jaffe et al., Phys. Rev. {\bf D40} (1989) 2777.

\bibitem{Nikif}  N.A. Nikiforov et al., Phys. Rev. {\bf C22} (1980) 700.

\bibitem{Cronin}
J.W.Cronin {\it et al.}, Phys.Rev. {\bf D11} (1975) 3105;\\
D.Antreasyan {\it et al.}, Phys. Rev. {\bf D19} (1979) 764.


 \bibitem{Protvino}

V.V.Abramov et al.,  Sov.\ J.\ Nucl.\ Phys.\ {\bf 41} (1985) 700.








\bibitem{Stavinsky} V.S. Stavinsky,
Physics of Elementary Particles and Atomic Nuclei
{\bf 10}, 949 (1979).


%



%

 \bibitem{QM99}
 {\it
 Proceedings of the 14th International Conference
 on Ultra-Relativistic Nucleus-Nucleus Collisions},
 Torino, Italy, 1999,  edited by L.Riccati et al.;
  Nucl. Phys.  {\bf A661}  (1999).

 \bibitem{QM01}
 {\it    Proceedings of the 15th International Conference
 on Ultra-Relativistic Nucleus-Nucleus Collisions},
 Long Island, New York, USA, 2001, edited by T.J.Hallman et al.;
  Nucl. Phys.  {\bf A698}  (2002).


 \bibitem{Zolin}  O.P. Gavrishchuk  et al.,
 Nucl. Phys. {\bf A523} (1991) 589.


\bibitem{ITEP}  S.V. Boyarinov   et al.,
 Sov.\ J.\ Nucl.\ Phys.\


\bibitem{Leksin}
 G.A. Leksin, Report No. ITEP-147, 1976 (unpublished) ;
 G.A. Leksin, in {\it Proceedings of the XVIII International
Conference  on High Energy Physics}, Tbilisi, Georgia, 1976,
 edited by N.N. Bogolubov {\it et al.} (JINR Report No. D1,2-10400,
 Tbilisi, 1977), p. A6-3.


 \bibitem{Strikman}  L.L. Frankfurt, M.I. Strikman,
 Phys. Rep. {\bf 160} (1988) 235.



 \bibitem{Bondarev}  V.K. Bondarev,
 Physics of Elementary Particles and Atomic Nuclei {\bf 28} (1997) 13.

 \bibitem{ABaldin}  A.A. Baldin, A.M. Baldin,
 Physics of Elementary Particles and Atomic Nuclei
 {\bf 29} (1998) 577.



%


%
%

%


%


%

\end{thebibliography}
\end{document}